\documentclass[aps,prl,twocolumn,letter]{revtex4-1}
\usepackage{graphicx}
\usepackage{float}
\usepackage{amsmath}
\usepackage{color}

\newcommand{\winding}[2]
{
\phi_{{#1},{#2}}
}

\newcommand{\distance}[2]
{
r_{{#1},{#2}}
}

\newcommand{\phiVar}
{
\sigma^2_\phi
}

\newcommand{\phiVarMicro}
{
\sigma^2_{\phi,0}
}

\newcommand{\rMean}
{
\langle r \rangle
}

\newcommand{\rVar}
{
\sigma_r^2
}

\begin{document}

\title{Atomistic study of macroscopic analogs to short chain molecules}
\date{\today}
\author{Kyle J. Welch, Clayton S. G. Kilmer, Eric I. Corwin}
\affiliation{Materials Science Institute and Department of Physics, University of Oregon, Eugene, Oregon 97403}

\date{\today}

\begin{abstract}

We use a bath of chaotic surface waves in water to mechanically and macroscopically mimic the thermal behavior of a short articulated chain with only nearest-neighbor interactions.  The chaotic waves provide isotropic and random agitation to which a temperature can be ascribed, allowing the chain to passively explore its degrees of freedom in analogy to thermal motion.  We track the chain in real time and infer end-to-end potentials using Boltzmann statistics.  We extrapolate our results, by using Monte Carlo simulations of self-avoiding polymers, to lengths not accessible in our system.  In the long chain limit we demonstrate universal scaling of the statistical parameters of all chains in agreement with well-known predictions for self-avoiding walks.  However, we find that the behavior of chains below a characteristic length scale is fundamentally different.  We find that short chains have much greater compressional stiffness than would be expected.  However, chains rapidly soften as length increases to meet with expected scalings.

\end{abstract}

\maketitle

Anyone who has ever put the wrong weight motor oil into a car engine can attest to the fact that the length of a chain molecule largely determines its mechanics \cite{fetters_connection_1994}.  A low-weight oil may be too thin and a high-weight oil too viscous for efficient operation.   Biopolymers such as DNA also evidence changing mechanical behavior with changing length.  Recent work shows that short strands are far more flexible than would be expected from simply scaling down the behavior of long strands \cite{vafabakhsh_extreme_2012}.  Understanding this scale dependence in polymers is crucial to creating new materials, as is being done with polymer thin films \cite{buchko_processing_1999,ouyang_programmable_2004} and so-called ``DNA origami" \cite{winfree_design_1998,rothemund_folding_2006}.  Polymer mechanics is well explored in the coarse-grained sense, with many established methods for direct measurement  \cite{smith_direct_1992,domke_measuring_1998,bustamante_ten_2003} and simulation \cite{grest_molecular_1986,honeycutt_general_1998,muller-plathe_coarse-graining_2002,fritz_coarse-grained_2009}.  However, there is scant experimental evidence directly relating whole polymer mechanical properties to behavior at the single bond level.  These questions have been addressed in macroscopic granular polymer studies \cite{ben-naim_knots_2001,zou_packing_2009} however such systems lack a thermodynamic temperature, muddying the link to true thermal systems.  Statistical physics predicts \cite{gennes_scaling_1979}, and empirical studies confirm \cite{jensenius_measuring_1997} that the end-to-end potential of a sufficiently long polymer chain is harmonic, regardless of the microscopic interactions.  Similarly, universalities are predicted for the scaling of statistical parameters (i.e., variance and mean) of both bond winding angle \cite{duplantier_winding-angle_1988} and linear dimension \cite{flory_configuration_1949,flory_principles_1953} in a polymer chain when considered as a self-avoiding walk (SAW).  While this is a powerful and robust result regarding the chain-scale mechanics, it does little to elucidate the behavior towards the monomer scale, as it does not address short chains.
\begin{figure}[th]
	\includegraphics[width=0.5\textwidth]{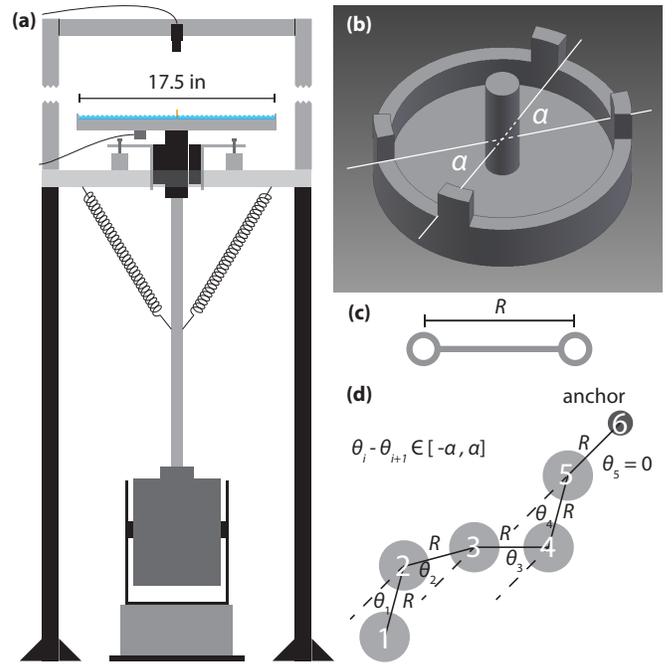}
	\caption{(Color online) (a) The shaker assembly.  (b) 3D schematic of one for the buoyant particles in the chain, showing the definition of restricting angle $\alpha$.  (c) A link showing fixed length $R$.  (d) A chain, demonstrating the measurement of link angle $\theta_i$.  Bulk rotations are eliminated from our coordinate system by measuring all $\theta_i$ off of the link most proximal to the anchor.  This allows us to work exclusively in terms of the relative coordinates of members of the chain.}
	\label{fig:diagrams}
\end{figure}

In this paper we build a macroscopic analog to polymer physics in a bottom-up fashion which allows us to observe not only chain-scale but also true monomer-scale behavior.  We find that short polymer chains exhibit behavior fundamentally different than that predicted for their long counterparts.  We use chaotic Faraday waves to create a macroscopic thermal-like bath with a well-defined temperature $\tau$, as we have previously described in reference \cite{welch_ballistic_2014}.  The waves are generated by vertically oscillating a circular aluminum dish with a diameter of 17.5 inches and a depth of 0.5 inches (Fig \ref{fig:diagrams}a).  The dish is mounted on a ceramic-coated aluminum shaft (diameter 2 inches), which passes through a 2 inch inner diameter air bushing (New Way Air Bearings).  This assembly is leveled by a stage mounted to the outside of the bushing mounting block.  The ceramic coated shaft is connected by a 24 inch length of 1 inch T-slot framing to an electromechanical shaker (Vibration Test Systems VTS-100) powered by a digital function generator (Stanford Research Systems Model DS335) through a voltage amplifier (Behringer Europower EP4000).  An accelerometer (CTC AC244-2D/010) is mounted to the bottom of the dish to measure stroke amplitude and frequency.

We 3D-print our macroscopic polymer analogs in a clear photopolymer resin using a Formlabs Form1.  Our macroscopic polymer consists of two basic elements: buoyant particles and links.  The particles are cylindrical boats (diameter 15mm, height 3mm) with centered masts  (diameter 2mm, height 7.5mm).  The links are rods with loops at each end, separated by 22mm on center, that slide over the masts to link the boats together in a chain (Fig. \ref{fig:diagrams}c).  Bond angle is restricted by stops on the gunwales of the boats (Fig. \ref{fig:diagrams}b) separated by an angular distance $\alpha$.  This chain is attached to a brass post set into the center of the aluminum dish using a link that is free to rotate about the anchor point.  This prevents the chain from interacting with the boundary of the dish, where the physics is fundamentally different, and allows us to gather data for arbitrarily long amounts of time, in our case 8 hours.  In this study we present data for: 1) a chain with $\alpha = \pi/4$ consisting of the post plus five other particles, five links, and 2) a chain with $\alpha = \pi/6$ consisting of the post plus eleven particles, and eleven links.

To facilitate automated tracking each boat has a circular well imprinted in the bottom with a diameter 1/2 that of the boat itself, uniformly filled with black oil paint.  The anchor post is capped with a black tracking point.  The positions of these black spots are tracked in real time at 30 frames per second (Pointgrey Flea3 USB camera with a Pentax C32500 KP lens, mounted 8 feet above the dish) using a radial symmetry algorithm \cite{parthasarathy_rapid_2012}.  To gather data on the dynamics of the chain we vertically oscillate our system with a frequency of 40Hz and an amplitude of 0.18mm peak-to-peak creating a temperature of about $1.3 \times 10^8$ J.  These parameters were chosen such that the Faraday waves would be well into the chaotic regime without causing the water to splash and capsize the boats.  We are limited in the length scales that we may probe in our system. Particles or chains smaller than the typical Faraday wavelength (on the order of millimeters) will not receive thermal-like agitation.  Chains longer than the diameter of our dish (44.5 cm) can interact with the side walls.  The dish size was chosen as to allow for the largest possible system while still remaining within the driving capability of our shaker.

\begin{figure}[th]
	\includegraphics[width=.5\textwidth]{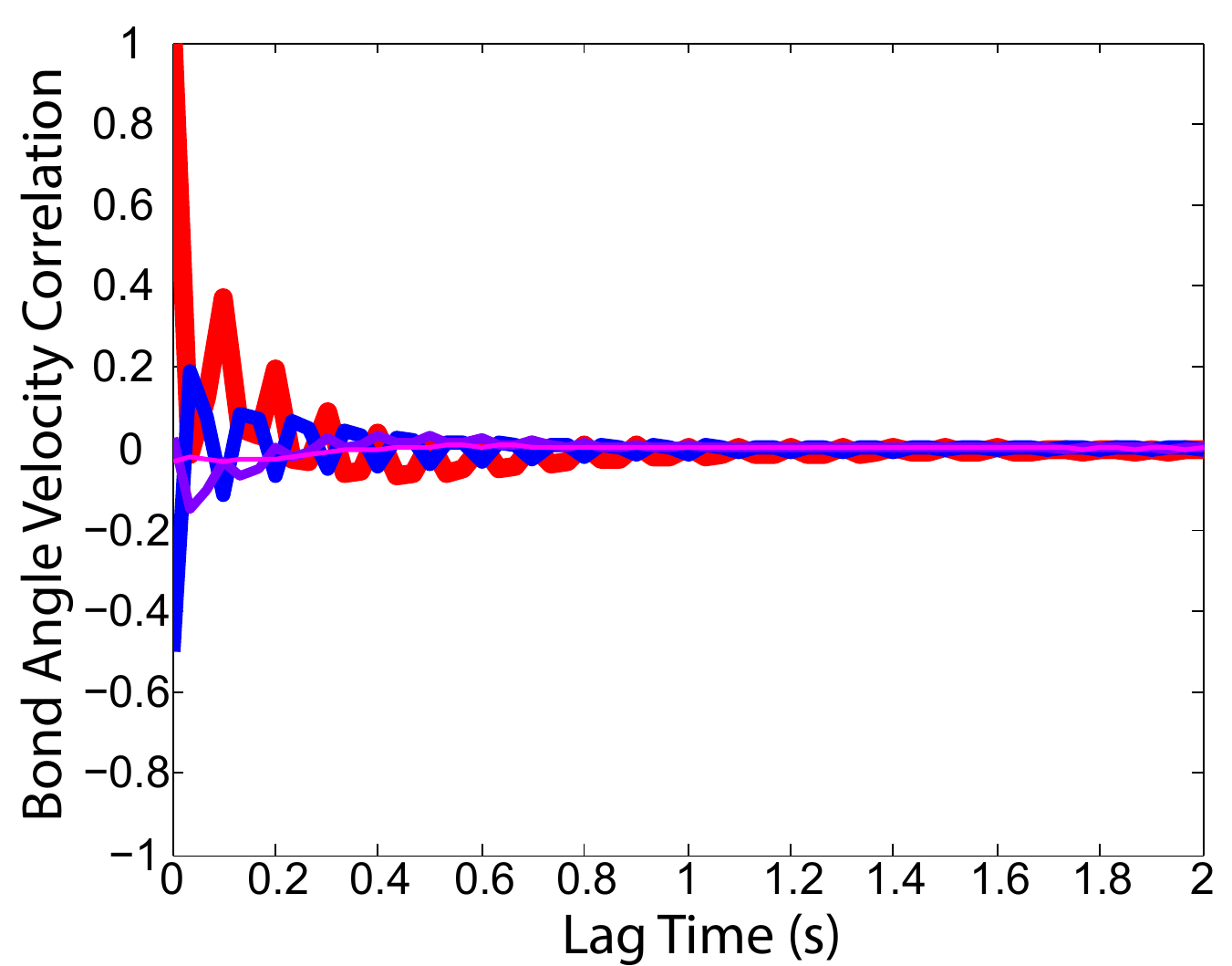}
	\caption{(Color online) Average bond angle velocity autocorrelation shown in red (thickest curve) for a chain with $\alpha = \pi/6$.  Nearest, second nearest, and third nearest bond angle velocity neighbor correlations shown in blue, purple, and magenta (second, third, and fourth thickest curves) respectively.  The nearest neighbor bond angle velocities are initially anticorrelated due to buckling in the interconnected chain, i.e. as one bond angle is kicked its neighbor will react in the opposite direction.  All correlations die off by $\sim0.3$s of lag time.  The 10Hz beat frequency results from the non-commensurate dish oscillation frequency (40Hz) and imaging frequency (30Hz).}
	\label{fig:correlation}
\end{figure}

In Figure \ref{fig:correlation} we demonstrate that the instantaneous bond angle velocities induced by the chaotic waves are uncorrelated in a chain of rigidly linked particles for times longer than the ballistic timescale (about 0.3 s).  The motion of a given bond angle is initially anticorrelated with its immediate neighbor due to the buckling of a chain of rigid links. As a bond angle is kicked in one direction, its neighbor must react in the opposite direction as a result of the constraints in the system.  More distant neighbors show weak correlations that die off quickly.  The 10Hz oscillation in correlation results from the beating between the dish oscillation frequency (40Hz) and the imaging frequency (30Hz).  The lack of correlation beyond that induced by buckling of the chain allows us to rule out the possibility of non-thermal correlations between particles.

To simulate the behavior of chains longer than experimentally accessible we employ a simple unbiased Monte Carlo algorithm to simulate each chain as a SAW. We begin by generating a random walk with $N$ bond angles where each step is of length $R$ and taken in a direction drawn from the distribution of bond angles 
\begin{equation}
p(\winding{i}{i+1}) \propto e^{-V_\phi(\winding{i}{i+1})/\tau} 
\end{equation}
created by a given potential $V_\phi$.  The random walk is then checked for overlaps of the boats and/or the links.  If there are any overlaps, the chain is thrown out.  If there are no overlaps then the chain is indeed a SAW and the end-to-end distance and winding angle are computed.  This process is repeated until the data for $10^6$ valid SAWs of length $N$ have been collected.

We follow the Flory convention \cite{flory_principles_1953}, which utilizes bond lengths and bond angles to describe a polymer in generalized nearest-neighbor coordinates (Fig. \ref{fig:diagrams}d).  To coarsen beyond nearest-neighbor coordinates, we represent the configuration of the chain in terms of radial distance $\distance{i}{j}$ between particles $i$ and $j$ and a winding angle between bonds $i$ and $j$ defined as
\begin{equation} \label{eq:relAngle}
\winding{i}{j} = \sum_{m=i}^{j-1} (\theta_m - \theta_{m+1}) ,
\end{equation}
where $\theta_i$ is the angle of the $i$th bond with respect to the link nearest the anchor point.  The use of a bond winding angle ensures that the angles we deal with are cumulative quantities and not bounded between $\pm \pi$.  Note that, since $n_{bonds} = n_{particles}-1$, the winding angle $\winding{i}{j}$ corresponds to the interparticle distance $\distance{i}{j+1}$.

\begin{figure*}
	\includegraphics[width=\textwidth]{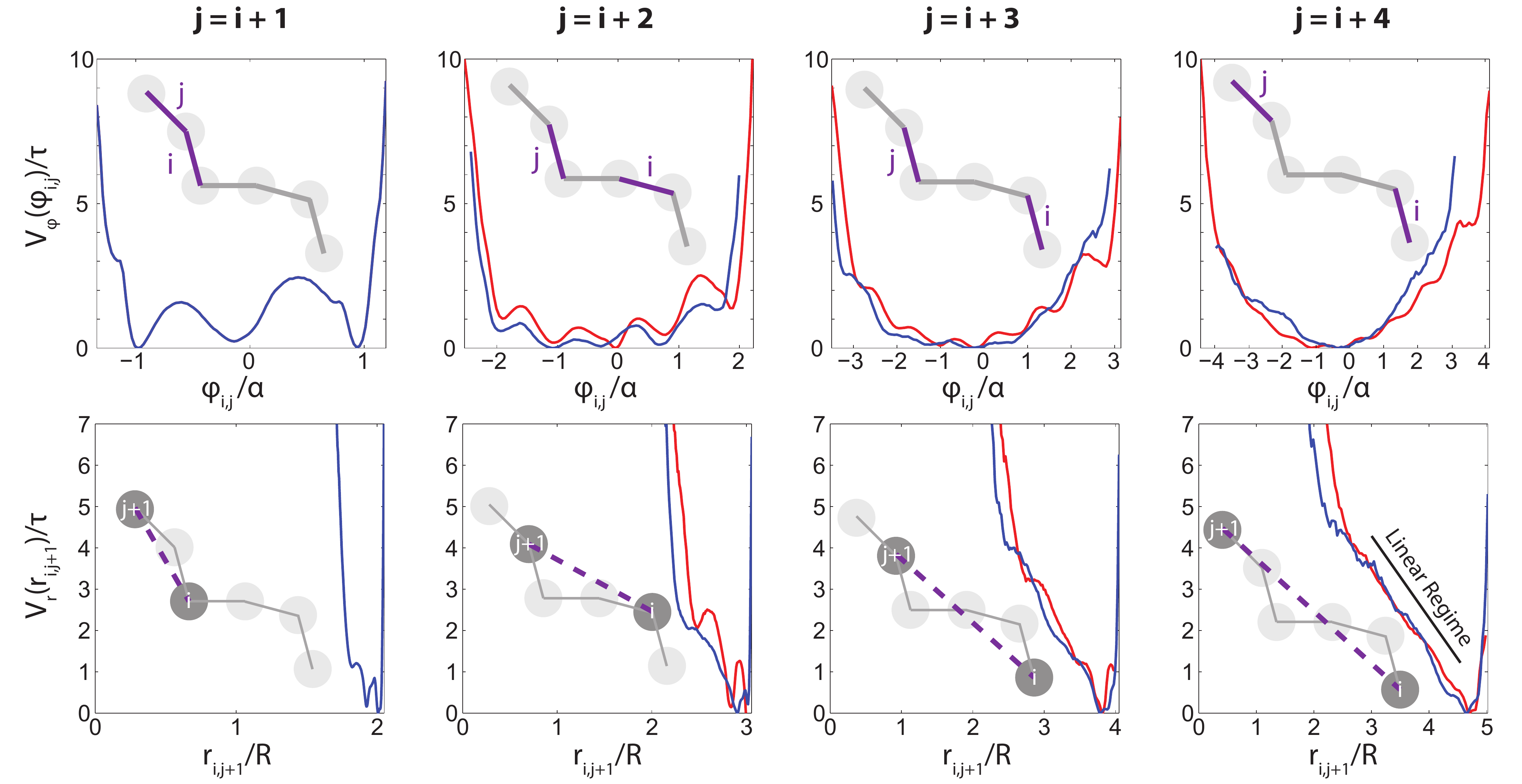}
	\caption{(Color online) Effective potentials for total bond winding angle $\winding{i}{j}$ (top row) and interparticle distance $r_{i,j+1}$ (bottom row) with example subchains shown as inset diagrams.  Potentials are determined from distributions using equations \ref{eq:boltzmannAngle} and \ref{eq:boltzmannR}.  The blue (dark) curves are calculated directly from empirical data and the red (light) curves are calculated using the distribution for $\winding{i}{i+1}$ to generate self-avoiding walks.}
	\label{fig:potentials}
\end{figure*}

We use Boltzmann statistics to measure effective potentials for the winding angle $V_{\phi}$ and the radial distance $V_r$.  We relate the probability distributions of winding angle $p(\winding{i}{j})$ and end-to-end distance $p(\distance{i}{j})$ to the underlying potentials scaled by the effective temperature $\tau$ \cite{welch_ballistic_2014} as 
\begin{equation} \label{eq:boltzmannAngle}
p(\winding{i}{j}) \propto e^{-V_\phi(\winding{i}{j})/\tau} 
\end{equation}
and
\begin{equation} \label{eq:boltzmannR}
p(\distance{i}{j}) \propto \distance{i}{j}e^{-V_r(\distance{i}{j})/\tau}. 
\end{equation}

We show the measured effective potentials averaged over all possible sub-chains of each given length for a system with $\alpha = \pi/4$ as blue (dark) curves in Figure \ref{fig:potentials}.  We present similar data for $\alpha = \pi/6$ in supplemental figures S1 and S2.  Note that the measured potentials for both systems reach beyond $\pm \alpha$.  A small amount of play was required to allow the links to slide without binding, creating a broadening of the potential.  Further, water bridges form where links contacts stops creating a more complicated potential energy landscape near the boundaries.  However, as we show in this paper, these details are immaterial as the variance of the nearest-neighbor bond angle distribution is all that matters in determining the physics of the system.

The winding angle potential $V_\phi$ is anharmonic for short chains but approaches a harmonic potential with increasing chain length, as predicted by a simple application of the renormalization group \cite{gennes_scaling_1979}.  The effective potential for end-to-end distance $V_r$, however, does not approach a harmonic within the length of chain empirically available to us for both the $\alpha = \pi/4$ and $\pi/6$ data.  This stands in contrast to the harmonic behavior expected for a freely-jointed chain, due to the fact that our chain has restrictions on bond angles and is by its nature non-intersecting.  Our data are much better to match to the worm-like chain model, showing qualitatively similar end-to-end distributions (and equivalent potentials) \cite{mehraeen_end--end_2008}.
\begin{figure}
	\includegraphics[width=.5\textwidth]{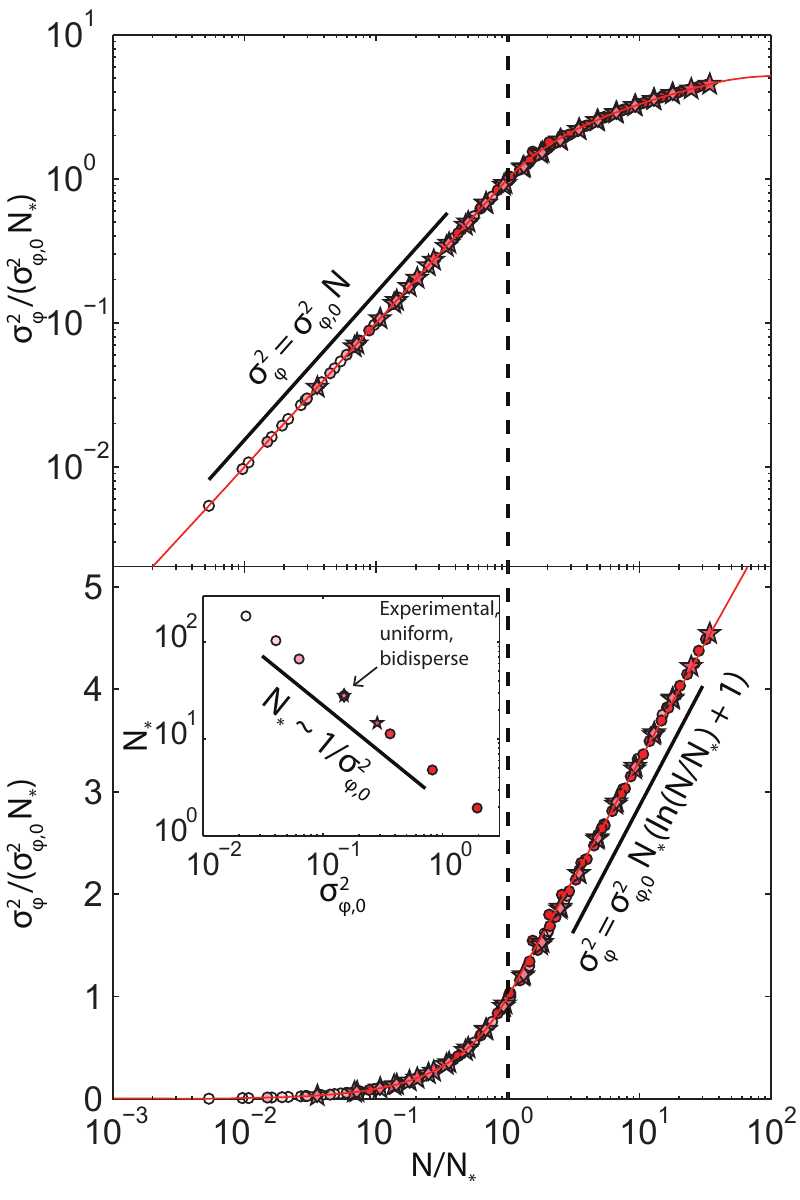}
	\caption{(Color online) Master curves for the scaling of winding angle variance $\phiVar$.  The data are plotted in both log-log (top) and semilog (bottom) to illustrate power law and logarithmic scaling regimes, respectively.  The shape of the markers indicates the probability distribution of nearest-neighbor link angles; stars for experimental, circles for uniform, and diamonds for bidisperse.  The marker color goes from light to dark as the variance of the nearest-neighbor probability distribution increases.  Equation \eqref{eq:thVarFit} is plotted over the data. (inset) The critical length $N_\ast$ is inversely proportional to $\phiVarMicro$, the variance of the nearest-neighbor link angle for a given chain.}
	\label{fig:thVar}
\end{figure}

In order to understand the behavior of these potentials for long chains we turn to simulations of our polymer.  The red (light) curves in figure \ref{fig:potentials} show the effective potentials for simulated chains of the same lengths as our experimental results.  The simulated data agrees with the empirical data, confirming the validity of our simulation approach.  

We extrapolate to long chains in order to determine the scaling of the statistical parameters of the distributions of end-to-end distance and end-to-end winding angle.  End-to-end quantities are denoted as coordinates without $i,j$ subscripts.  Figure \ref{fig:thVar} shows the scaling of the variance of end-to-end winding angle $\sigma^2_\phi$ with $N$, the number of defined bond angles.  We find that for sufficiently short chains, $\phiVar$ is linear in $N$.  This is identical to the scaling expected for a simple random walk because at small scales self-intersections are unlikely due to the restriction of bond angles.  For sufficiently long chains we find that $\phiVar$ exhibits a logarithmic scaling, as predicted by \cite{duplantier_winding-angle_1988}.  To determine the critical $N$ at which chains transition between ``short'' and ``long'' we fit the data to the piecewise function
\begin{equation}
\phiVar = \begin{cases} 
	    \phiVarMicro N & N<N_\ast \\ 
	    \phiVarMicro N_\ast(\ln(N/N_\ast) + 1) & N>N_\ast 
	  \end{cases}.
\label{eq:thVarFit}
\end{equation}
where $\phiVarMicro$ is the already known variance of the nearest-neighbor link angle.  The fit yields the value of $N_*$, which we find to be inversely proportional to $\phiVarMicro$ (Figure \ref{fig:thVar} (inset)) with a phenomenological prefactor of order 1 that is likely determined by the relative size of the particles to the links.  $N_*$ plays a similar role to a Kuhn length in a real polymer.  We can collapse all of our data onto a master curve when scaled so that
\begin{equation}
N \to \frac{N}{N_\ast} \: , \: \phiVar \to \frac{\phiVar}{\phiVarMicro N_\ast}.
\end{equation}
We use three different families of bond angle distributions: 1) The experimentally measured distributions from our $\alpha = \pi/4$ and $\alpha = \pi/6$ systems, the corresponding potentials for which are shown in the top left of Figure \ref{fig:potentials} and S1, respectively. 2) A uniform distribution of bond angles between $\pm \alpha$. 3) A bidisperse distribution of bond angles with value of $\pm \alpha$.  We present simulational data for SAWs from uniform and bidisperse distributions tuned to have the same $\phiVarMicro$ as the $\alpha = \pi/6$ empirical data and find that all three associated scalings have precisely the same $N_\ast$.  Thus we see that the full physics of the system is controlled by the variance of the nearest-neighbor angle distribution.

\begin{figure}
	\includegraphics[width=.5\textwidth]{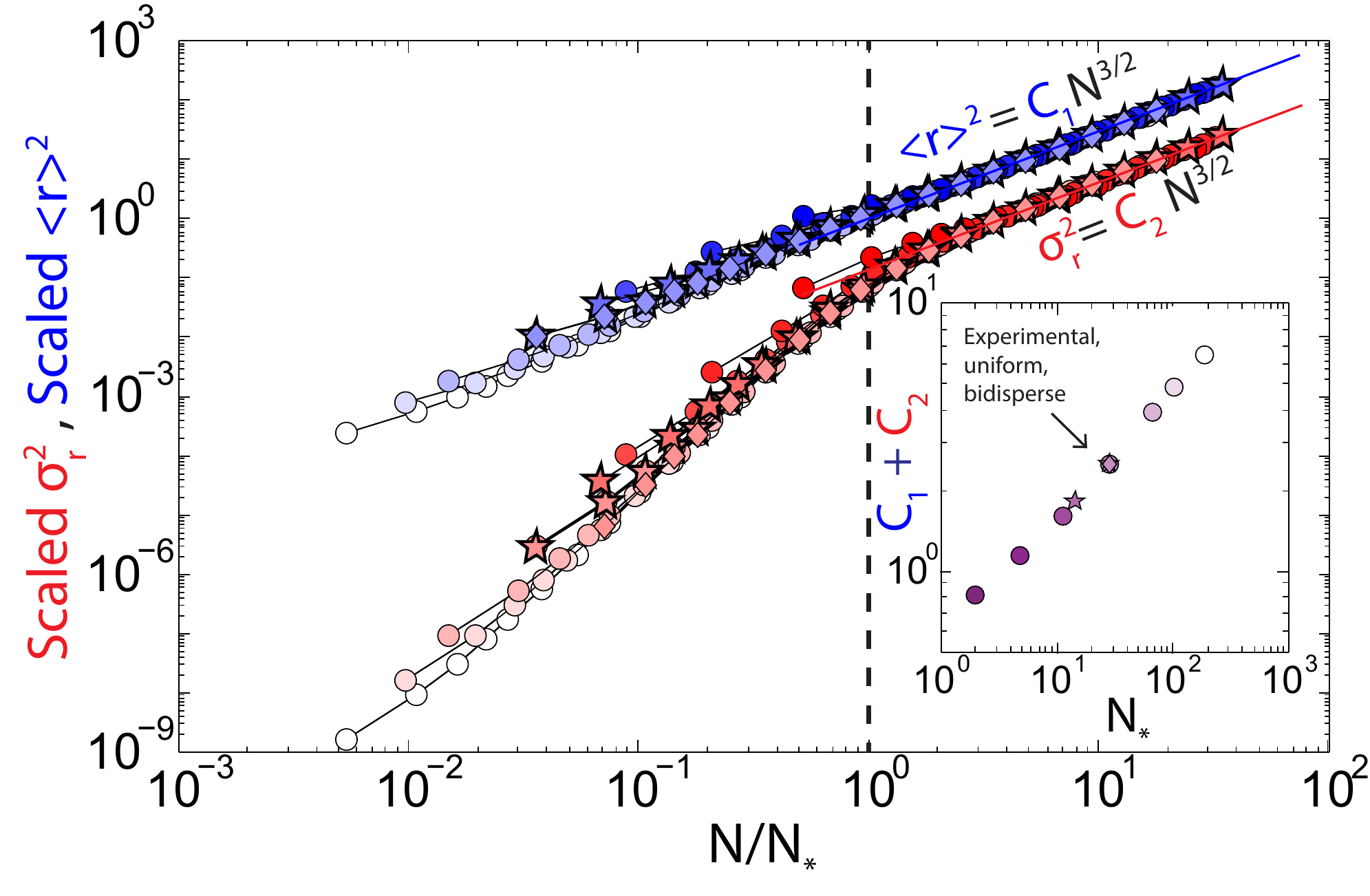}
	\caption{(Color online) Scaling of variance in end-to-end distance $\rVar$ (lower curve) and squared mean of end-to-end distance $\rMean^2$ (upper curve).  The shape and color of the markers indicates the probability distribution of nearest-neighbor bond angles as in figure \ref{fig:thVar}.  Both curves are scaled by a factor of $(C_1 + C_2)N_\ast^{3/2}$, where $N_\ast$ is the same critical length from figure \ref{fig:thVar}, and collapse to $N^{3/2}$ power laws for $N>>N_\ast$.  (inset) The sum of scaling coefficients $C_1$ and $C_2$ plotted against critical length $N_\ast$.}
	\label{fig:rMeanVar}
\end{figure}

Figure \ref{fig:rMeanVar} shows the scaling of the squared mean $\rMean^2$ and the variance $\rVar$ as a function of $N$. We see that all curves, for sufficiently large $N$, go to a power law consistent with the Flory prediction \cite{flory_configuration_1949,flory_principles_1953}, where
\begin{equation}
\rMean \sim N^{\nu} \: , \: \rVar \sim N^{2\nu} \: , \: \nu=\frac{3}{4}
\end{equation}
in two dimensions. We collapse all of our data onto a master curve for $N>>N_\ast$ by scaling $N \to N/N_\ast$ and
\begin{equation}
\rMean^2 \to \frac{\rMean^2}{(C_1 + C_2)N_\ast^{3/2}} \: , \: \rVar \to \frac{\rVar}{(C_1 + C_2)N_\ast^{3/2}}
\end{equation}
where the constants $C_1$ and $C_2$ are fitted prefactors (figure \ref{fig:rMeanVar}).  This scaling was selected such that $\langle r^2 \rangle = \rMean^2 + \rVar = 1$ at $N = N_\ast$, preserving relative magnitude of both statistical parameters.

Our simulations show that the for lengths much longer than experimentally accessible the radial potentials eventually become harmonic as expected (Figure S3).  However, for short and intermediate lengths the curves show non-trivial behavior.  The compression stiffness of the polymer, which is related inversely to the variance in end-to-end distance, is far higher in short chains than would be predicted by the Flory power law.  As these chains grow, the stiffness necessarily softens very rapidly in order to meet the long chain limit and match expectations above the crossover from ``short" to ``long."

Our system shows clear hallmarks of semiflexibility \cite{wilhelm_radial_1996,mehraeen_end--end_2008}, with the end-to-end potential transitioning from a rigid state in which the equilibrium position is near full extension of the chain, to a flexible state with an equilibrium position near zero (Figure S3).  We see further evidence of semiflexibility in the existence of a linear regime in our experimentally determined potentials (Figure \ref{fig:potentials}, lower right).  In the intermediate lengths between the two states, the potential temporarily becomes wide and flat before becoming harmonic.  This intermediate-scale widening explains the rapid scaling of the variance at these lengths.  This roll-over phenomenon is a result of our bonds being treated as rigid rods.  This assumption is valid as considering only torsional degrees of freedom is a standard in the field, consistent with the rotational isomeric state model of polymers \cite{flory_statistical_1969,mattice_conformational_1994}.

In this study we have presented a simple mechanical analog to short chain molecules.  We see that the effective interaction potential for winding angle quickly becomes harmonic as chain length increases, despite the decidedly non-harmonic nearest-neighbor interaction.  Though the effective potential for end-to-end distance does not become harmonic over the length of our empirical chain, we are able to extrapolate for considerably longer chains and show that it eventually assumes the familiar Hookean form.  We show that our system may be treated as a self-avoiding walk with scalings that are consistent with \cite{duplantier_winding-angle_1988} and  \cite{flory_configuration_1949,flory_principles_1953} in the long-chain limit and that the full character of these scalings is determined by the width of the nearest-neighbor bond angle distribution.  We show that short- and intermediate-length chains exhibit unconventional behavior, a result likely connected the relative rigidity of bond lengths.  Ultimately, we demonstrate that over the full range of lengths studied all of the physics is controlled by the variance of the nearest neighbor bond-angle distribution.

Our simulation method accurately predicts the behavior of our mechanical system at the short scale as well as the well-known behavior of polymers in the long-chain limit.  Therefore, our mechanical analog is consistent with polymer physics and may be used to speak to the non-universal mechanics of short chain molecules.  Our macroscopic, mechanical system is uniquely suited to passively study these lengths as we can build the physics from the bottom up all while operating at easily observable length and time scales.  It is remarkable that such a simple macroscopic analog can point to a better understanding of the constitutive mechanics of microscopic polymers.  Beyond the study of polymers, the relative link angles in our chain can be mapped as spins onto a 1D lattice, so this experiment can be viewed as a simple mechanical simulation of a 1D X-Y model with peculiar nearest neighbor interactions.

%%%%%%%%%%%%%%%%%%%%%%%%%%%%%%%%%%%%%%%%%%%%%%%%%%%%%%%

\section{Acknowledgements}

We thank Raghuveer Parthasarathy for the radial symmetry tracking algorithm, John Royer for helpful discussions, and Oliva Carey-De La Torre and Cody Hill for initial machine work.  We thank the NSF for support under Career Award No. DMR-1255370.

\bibliography{MacroPolymerBib}

\end{document}